\documentclass{jaa}
\usepackage{natbib}
\usepackage{url}
\usepackage{graphicx}
\usepackage[multiple]{footmisc}
\usepackage{hyperref}

\newcommand{\photu}{photon units}

\newcommand{\galex}{{\it GALEX}}

\newcommand{\chisq}{$\chi^{2}$}
\newcommand{\ebv}{E(B~-~V)}

\providecommand{\apjl}{Astrophys.\ J.\ Lett.}
\providecommand{\apjs}{Astrophys.\ J.\ Suppl.}
\providecommand{\aap}{Astron.\ Astrophys.}
\providecommand{\mnras}{Mon.\ Not.\ R.\ Astron.\ Soc.}
\providecommand{\apj}{Astrophys.\ J.}
\providecommand{\araa}{Annu.\ Rev.\ Astron.\ Astrophys.}
\providecommand{\pasp}{Publ.\ Astron.\ Soc.\ Pac.}
\providecommand{\apss}{Astrophys.\ Space Sci.}
\providecommand{\prd}{Phys.\ Rev.\ D}
\providecommand{\aj}{Astron.\ J.}

\begin{document}

\title{Modeling the Cosmic Ultraviolet Background at the North Galactic Pole}

\author{Jayant Murthy}
\affilOne{\textsuperscript{1}Indian Institute of Astrophysics, Bengaluru 560 034, India.\\}

\twocolumn[{

\maketitle

\corres{jmurthy@yahoo.com}

\msinfo{4 May, 2025}{13 June, 2025}

\begin{abstract}

I explore models of the dust-scattered component of the Cosmic Ultraviolet Background (CUVB) at the North Galactic Pole (NGP) in order to develop a framework for calculating the dust-scattered light as a function of the optical depths. As expected, I find that the dust-scattered emission scales linearly with reddening up to $E(B-V) \approx 0.1$\ mag and derive a parametric model for this dependence. 
I have applied these models to fit the far-ultraviolet (1350--1800 \AA) observations from the \textit{Galaxy Evolution Explorer (GALEX)} finding that the optical constants of the interstellar dust grains --- albedo ($a$) and phase function asymmetry factor ($g$) --- are consistent with predictions from the Astrodust model ($a = 0.33$, $g = 0.68$). I detect an isotropic offset of $267 \pm 7$ ph cm$^{-2}$ s$^{-1}$ sr$^{-1}$ \AA$^{-1}$, half of which remains unaccounted for by known Galactic or extragalactic sources. 

I will now extend my analysis to wider sky regions with the goal of generating high-resolution extinction maps.
\end{abstract}

}]


\doinum{12.3456/s78910-011-012-3}
\artcitid{\#\#\#\#}
\volnum{000}
\year{0000}
\pgrange{1--8}
\setcounter{page}{1}
\lp{1}

\section{Introduction} \label{sec:intro}

The cosmic ultraviolet background (CUVB) is, by definition, comprised of the diffuse Galactic light (DGL) and the extragalactic background (EBL) \citep[reviewed by][]{Bowyer1991, Henry1991, Murthyreview2009}. The DGL dominates the CUVB at low Galactic latitudes, where it is largely made up of dust-scattered light that is correlated with the reddening at low optical depths ($\tau$) but saturates when $\tau > 1$ \citep{Murthy_galex_data2010, Seon2011, Hamden2013, Murthy2014apj}. Other contributors to the DGL include molecular hydrogen fluorescence \citep{Martin1990_h2, Jo2017_h2} excited by ultraviolet (UV) photons from the interstellar radiation field (ISRF); emission from highly ionized lines \citep{Martin1990_lines, Shelton2001}; two-photon emission \citep{Reynolds1990, Kulkarni2022}; and possible exotic contributors \citep{PorrasBedmar2024_axion} and contribute more of the DGL at high Galactic latitudes where there is little dust. The EBL contributes about 100 \photu\footnote{photons cm$^{-2}$ \AA$^{-1}$ s$^{-1}$ sr$^{-1}$.}\footnote{ 50.35 photon units = 1 nW m$^{-2}$ sr$^{-1}$ at any wavelength.}, or about a third of the total diffuse background, near the Galactic poles and is largely due to the light of unresolved galaxies \citep{Driver2016}.

\begin{table*}[t]
\centering
\caption{Polar Observations
\label{tab:polar}}
\begin{tabular}{cccc}
\hline\hline
References & Wavelength (\AA) & Zero-Offset$^{a}$ & Instrument \\
\hline
\citet{Henry_ngp1978} & 1180 -- 1680 & 250 & Apollo 17 \\
\citet{Anderson1979} & 1230 -- 1680 & $285 \pm 32$ & Rocket \\
\citet{Feldman_hotgas1981} & 1200 -- 1670 & 150 $\pm$ 50 & Rocket \\
\citet{Holberg1986} & 1100 & $< 200$ & Voyager \\
\cite{Onaka1991} & 1500 & 200 -- 300 & Rocket \\
\cite{Henry1993} & 1500 & 300 $\pm$ 100 & UVX \\
\cite{Witt1994} & 1500 & $300 \pm 80$ & DE-1 \\
\cite{Witt1997} & 1400 -- 1800 & $160 \pm 50$ & FAUST \\
\citet{Murthy_voy} & 1100 & $<30$ & Voyager \\
\cite{Schiminovich2001} & 1740 & $200 \pm 100$ & NUVIEWS \\
\cite{Puthiyaveettil2010} & 1400 -- 1900 & 500$^*$ & DE-1 \\
\cite{Hamden2013} & 1565 & 300 PU & Galex \\
\citet{Akshaya2018} & 1565 & $288 \pm 2$ & GALEX NGP\\
                    & 2365 & $531 \pm 2$ & \\
 \citet{Akshaya2019} & 1565 & $240 \pm 18$ & GALEX \\
                    & 2365 & $394 \pm 37$ & \\
\hline
\multicolumn{4}{l}{$^{a}$\photu}
\end{tabular}
\end{table*}

Characterizing the different phases of the CUVB is best done near the Galactic poles where the dust contribution is minimized and, as a consequence, there have been several observations of the CUVB near the Galactic Poles, almost all around the North Galactic Pole (NGP). These studies (Table \ref{tab:polar}) have invariably assumed that the dust-scattered light is linearly correlated with the reddening, estimated from either the gas column density \citep{Lenz2017} or the thermal emission from interstellar dust \citep{Schlegel1998, PlanckDust2016}, and have fit the observations with a straight-line fit. The slope of the linear fit represents the dust-scattered light and the y-intercept is referred to as the offset and is 200 -- 300 \photu\ at 1500 \AA, of which 65 -- 80 \photu\ is attributed to galaxies \citep{Driver2016}. Although some part of the remainder may be attributed to emission from halo gas, much of it cannot be accounted for, as yet, by known Galactic or extragalactic sources \citep{Henry2015, Chiang2019, Murthy2025_alice}.

In principle, the dust-scattered light should provide an estimator of interstellar extinction at a greater spatial resolution than possible from observations of the thermal emission of dust grains \citep{Boissier2015}. However, the amount of scattered light is dependent on the relative geometry of the dust and the stars \citep{Witt1997, Seon2011} and the local properties of the dust, which may vary across the sky \citep{Ysard2015}. A complete modeling of the dust-scattered light will necessarily involve a multiple scattering approach \citep[eg.][]{Murthy_dustmodel2016} but these are expensive in terms of computer resources. I have developed a parametrization of the scattering as a function of the optical depth that I apply to \textit{Galaxy Evolution Explorer} (\textit{GALEX}) observations of the CUVB in the far ultraviolet (FUV: 1350 -- 1800 \AA). In this work, I will focus on the development of the algorithm and the methodology.

\section{Modeling}

I have used the Monte Carlo model described by \citet{Murthy_dustmodel2016} and updated by \citet{Akshaya2019} to calculate the dust scattering as a function of optical depth at the NGP. The Galaxy is divided into bins and filled with dust using the 3-dimensional dust distribution of \citet{Green2019}. Bins that were not covered by the Green dust observations were filled using a dust distribution falling off from the Galactic Plane with a scale height of 125 pc \citep{Marshall2006} and a cumulative reddening to match the \ebv\ from \citet{Schlegel1998}. I assumed a cavity of radius 50 pc around the Sun \citep{Welsh2010}. In practice, the scattered radiation is relatively independent of the details of the dust distribution.

The interstellar radiation field in the UV is dominated by a small number of O and B stars whose spectral types and positions can be read from a catalog \citep{Henry_radiation_field1977, Murthy_model1995}. I have used stars from the Hipparcos catalog \citep{Perryman1997}, and modeled the emission from each star using theoretical spectra from \citet{Castelli2004}. Individual photons from each star were followed throughout their path until they interacted with a dust particle, at which time they were scattered using the Henyey-Greenstein scattering function \citep{Henyey1941}. Grain cross-sections as a function of wavelength were from \citet{Draine_scat2003} and are observationally determined from the interstellar extinction curve. The final result was a map of the dust-scattered radiation over the entire sky as a function of the albedo ($a$) and phase function asymmetry factor ($g$) of the interstellar dust grains. These maps have been discussed by \citet{Murthy_dustmodel2016} and are available from \url{https://doi.org/10.5281/zenodo.5337045}.

\begin{figure*}
    \includegraphics[width=3.5in]{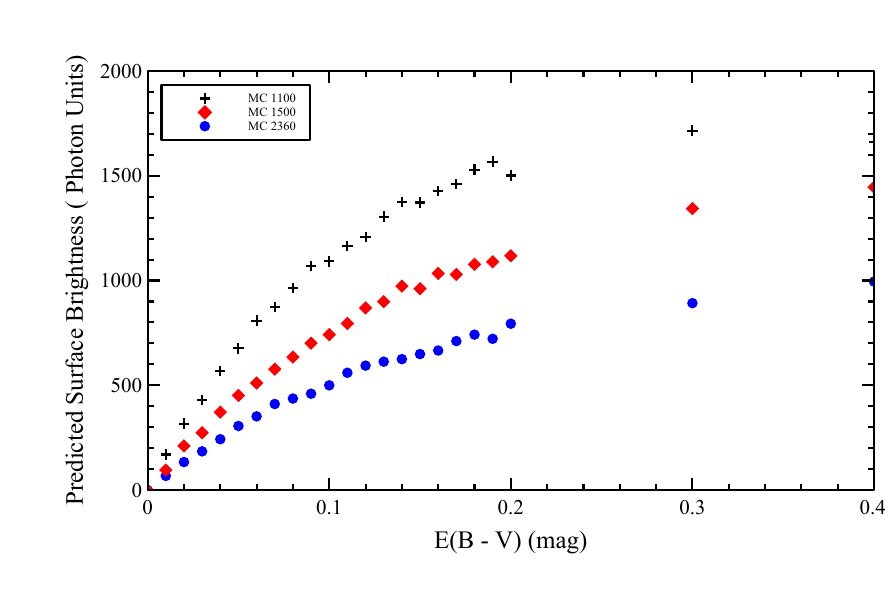}
    \includegraphics[width=3.5in]{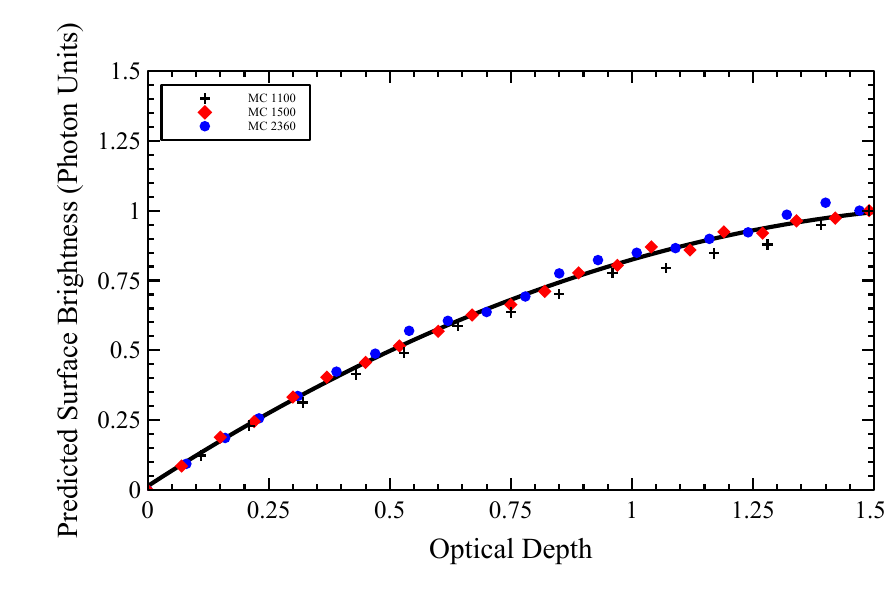}

    \caption{Model predictions for the dust-scattered light at the NGP as a function of \ebv\ on the left. I have fixed $a = 0.4$ and $g = 0.4$ and plotted for wavelengths of 1100 \AA\ (black), 1500 \AA\ (red), and 2360 \AA\ (blue). I have normalized the predictions by the ISRF and plotted them as a function of optical depth ($\tau$) in the right. The solid line shows the best-fit quadratic of Eq. \ref{eq:par}.} 
    \label{fig:ebv_range}
\end{figure*}

\begin{table*}[t]
\caption{Bright Stars
\label{tab:brightstars}}
\begin{tabular}{lcllcccll}
\hline\hline
HIP & Sp. Type & M$_{V}$ & GL & GB & Dist. &\multicolumn{3}{c}{Percentage of total}\\
& & mag &   & & pc & 1100 \AA & 1500 \AA & 2360 \AA\\
\hline
 60718 & B0.5IV & 0.6 & 300.1 & -0.4 & 98 & 6.6 & 4.4 & 4.1 \\
 68702 & B1III & 0.6 & 311.8 & 1.3 & 161 & 5.7 & 4.3 & 3.9 \\
 62434 & B0.5III & 1.3 & 302.5 & 3.2 & 108 & 4.1 & 2.7 & 2.2 \\
 26727 & O9.5Ib SB & 1.7 & 206.5 & -16.6 & 250 & 3.5 & 2.8 & 2.3 \\
 25930 & O9.5II & 2.2 & 203.9 & -17.7 & 280 & 2.1 & 1.7 & 1.3 \\
\hline
\end{tabular}
\end{table*}

\begin{table}[t]
\caption{Scattering Parameters
\label{tab:parameters}}
\begin{tabular}{llll}
\hline\hline
$a$ & $g$ & P0 & P1 \\
\hline
0.10 & 0.00 &    592.02527 &   -240.33853\\
0.10 & 0.01 &    599.57794 &   -246.27243\\
0.10 & 0.02 &    588.18060 &   -246.43423\\
0.10 & 0.03 &    610.61664 &   -274.55441\\
0.10 & 0.04 &    557.15161 &   -218.35941\\
0.10 & 0.05 &    591.68390 &   -274.35117\\
0.10 & 0.06 &    558.51215 &   -239.84650\\
0.10 & 0.07 &    548.86615 &   -213.60622\\
0.10 & 0.08 &    523.98383 &   -196.89223\\
0.10 & 0.09 &    532.59863 &   -202.71005\\
0.10 & 0.10 &    536.99762 &   -244.56952\\
\hline
\end{tabular}
\end{table}

Monte Carlo programs are statistical in nature and cannot be used to probe the fine-scale structure of the dust because they take too much time. I have explored parameterizations of the scattering that may be applied to more general solutions. I begin by calculating the predicted scattered radiation as a function of the optical constants and the total reddening at the NGP ($b > 80^{\circ}$) by fixing the \ebv\ in every line-of-sight above $70^{\circ}$ and running the Monte Carlo programs for a total of $10^{8}$ photons. As \citet{Jura1979} noted, the dust-scattered UV light at the NGP is primarily due to radiation from stars in the Galactic Plane back-scattered by interstellar dust. Further, because there are so few UV bright stars in the sky, 22\% of the dust-scattered emission at the NGP at 1100 \AA\ is due to only five stars (Table \ref{tab:brightstars}), with these stars contributing 16\% and 14\% to the emission at 1500 and 2360 \AA, respectively. 

I found that the scattering is approximately linear up to a reddening of about 0.1 mag, corresponding to an optical depth of 1.1 at 1100 \AA, and 0.8 at 1500 and 2360 \AA\ (Fig. \ref{fig:ebv_range}). As mentioned above, I have used \citet{Draine_scat2003} for the optical depth as a function of wavelength and plotted the scattering versus the optical depth in the right panel of Fig. \ref{fig:ebv_range}, after normalizing by the strength of the ISRF at each wavelength. The dust-scattered light for optical depths less than 1.5 was well fit by:
\begin{equation}
\label{eq:par}
    S(a, g,\tau) = P0(a, g)*\tau + P1(a, g)*\tau^{2}
\end{equation}
with a minimum $\chi^{2}$ of 0.859 (solid line in the right side of the figure). S is the predicted surface brightness at the NGP, $\tau$ is the optical depth, and P0 and P1 are the coefficients for the quadratic fit. Note that this relation breaks down for optical depths much greater than 1 because of saturation effects within the scattering medium.  A portion of the table is listed in Table \ref{tab:parameters} with the full comma-separated table available at \url{doi:10.5281/zenodo.15295073}.

I now have a prescription to fit the spectrum of the dust-scattered light at any location in the Galaxy where $\tau < 1$, if I account for the different ISRF and relative geometry. As pointed out earlier, it is too expensive to run Monte Carlo models for individual locations and I have therefore used a single-scattering model for each location to calculate the ISRF as a function of the optical constants and look direction, with the same dust model as described above.

\section{Fitting the Data}

\begin{figure*}
    \includegraphics[width=7in]{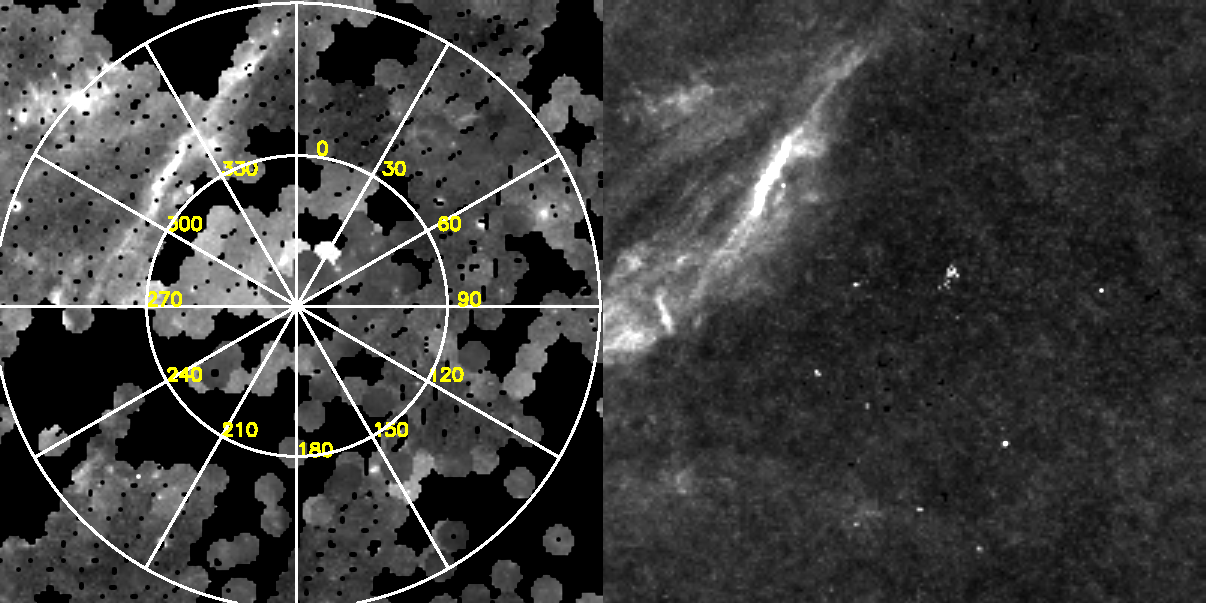}
    \caption{\galex\ FUV (left) and Planck \ebv\ maps (right) of the NGP at the same scale and projection. The coordinate scale is plotted on the FUV image with 80 and 85$^{\circ}$ lines shown. The prominent filaments in the FUV and \ebv\ maps are clouds first identified in polarization maps by \citet{Markkanen1979}. The black areas in the UV maps are regions that were not observed by \galex.} 
    \label{fig:data_images}
\end{figure*}

As an example of the fitting procedure, I model FUV (1300 -- 1800 \AA) observations at the NGP ($b > 80^{\circ}$) made with \galex\ \citep{Martin2005, Morrissey2007}. \citet{Murthy2014apj} extracted the diffuse background at a resolution of $15^{\prime\prime}$ from the original observations which I have rebinned into $6^{\prime}$ bins (Fig. \ref{fig:data_images}). I used the Planck dust map \citep{PlanckDust2016} for the reddening, shown on the right of the figure. \citet{Murthy2025_alice} showed that there was no emission from the interplanetary medium in the \galex\ FUV passband and, therefore, the \galex\ FUV surface brightness is a good measure of the CUVB at 1500 \AA.

\begin{figure}
    \includegraphics[width=3.5in]{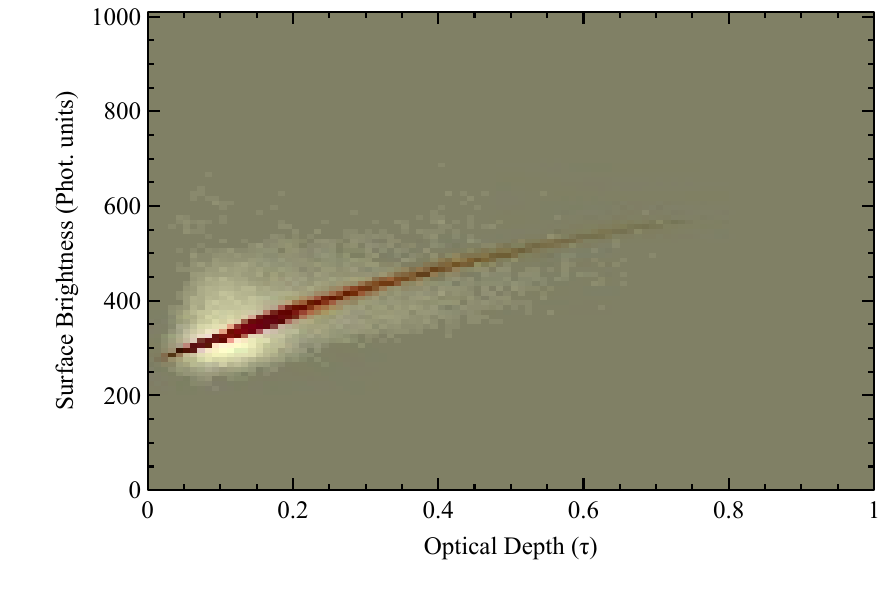}
    \caption{Density maps of data (grey/white) and the models.} 
    \label{fig:data_model}
\end{figure}

\begin{figure}
    \includegraphics[width=3.5in]{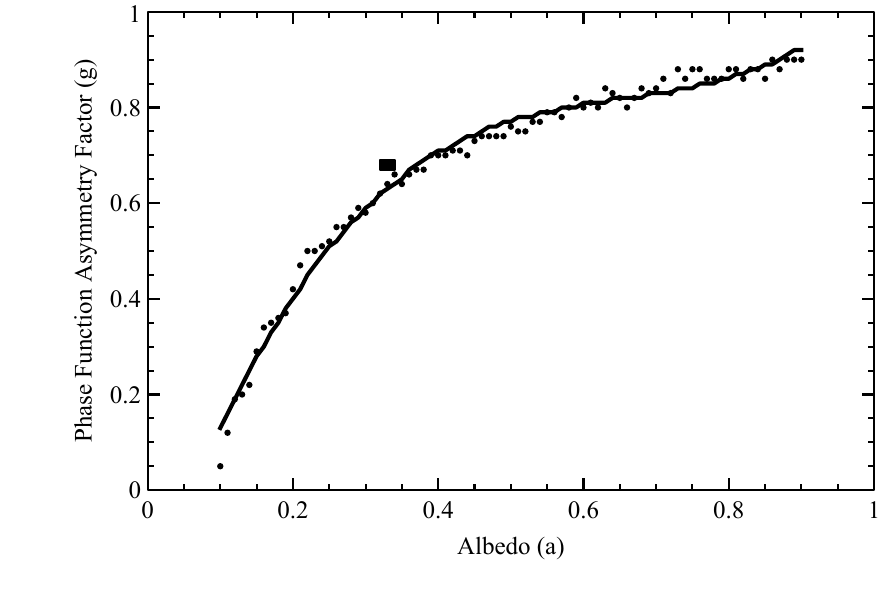}
    \caption{Allowed values for $a$ and $g$ for the \galex\ FUV. The black box shows the predicted values from \citet{Hensley2023}. The solid line is an empirical fit to the data (Eq. \ref{eq:ag}).} 
    \label{fig:plot_ag}
\end{figure}

The CUVB at high Galactic latitudes is the sum of the dust-scattered light and the extincted offset as represented by the following equation:
\begin{equation}
    CUVB = C(a, g)*S(a, g, \tau) + O * exp(-\tau).
\end{equation}
CUVB is the observed surface brightness in the FUV and NUV, S is the predicted dust-scattered surface brightness at the NGP (Eq. \ref{eq:par}), O is the offset, and $\tau$ is the optical depth at the given wavelength. C is the scaling for the ISRF and was calculated for each line of sight using the single-scattering model described above. The best fit to the data is shown in Fig. \ref{fig:data_model} with a minimum \chisq\ of 1.4. However, the error analysis is non-trivial because of the uncertainties in the Planck reddening ($\approx 5$ millimagnitudes), the FUV ($\approx 20$ \photu), and the model uncertainties (estimated to be $\approx 30$ \photu) and will be deferred to a future paper.

In this work, I focus on deriving $a$ and $g$ and the offset. with the allowed $a$ and $g$ plotted in Fig. \ref{fig:plot_ag}.  As \citet{Mathis2002} pointed out, there is a degeneracy between $a$ and $g$ because the radiation field is largely uniform over the NGP and I can fit the data equally well for any values of $a$ and $g$ where: \begin{equation}\label{eq:ag}
    g = -0.262 + 4.491*a - 6.512*a^2 + 3.318*a^3.
\end{equation}
This fit is shown in Fig. \ref{fig:plot_ag} with the predicted $a$ and $g$ of the Astrodust model \citep{Hensley2023} shown as a dark square. The Astrodust model assumes composite dust grains that are a mix of amorphous silicates and hydrocarbons. Note that the derived curve is purely empirical with no physical justification and is provided only for mathematical simplicity.

The offsets are independent of $a$ and $g$ and I find them to be $267 \pm 7$ \photu, agreeing well with earlier determinations (Table \ref{tab:polar}). As discussed by \citet {Murthy2025_alice}, approximately half of the signal may be explained by known sources.

\section{Conclusions and Future Work}

I have developed a methodology to calculate the dust-scattered light for optical depths less than unity ($\tau < 1$) and applied it to observations of the diffuse light in the NGP. Because the interstellar radiation field is uniform over the NGP, the optical constants are degenerate and I can only say that they are consistent with the model values of $a = 0.33$ and $g = 0.68$ from the Astrodust model \citep{Hensley2023}. The offset represents that part of the CUVB that is isotropic, and I have set a firm detection of $267 \pm 7$ \photu, of which about half can be explained through known Galactic and extragalactic sources.

I plan to apply these models to larger areas of the sky where there may be greater variation in the radiation field. I have shown that the expected dust scattering is independent of the optical constants over the NGP and is directly dependent on the amount of dust in the line of sight. This allows for the determination of an extinction map over the NGP to a sensitivity and spatial resolution unattainable from other methods and may be extended to other regions, as long as the optical depth is low.

\section*{Acknowledgments}

Some of the data presented in this paper were obtained from the Mikulski Archive for Space Telescopes (MAST). STScI is operated by the Association of Universities for Research in Astronomy, Inc., under NASA contract NAS5-26555. Support for MAST for non-HST data is provided by the NASA Office of Space Science via grant NNX13AC07G and by other grants and contracts. This research has made use of the SIMBAD database, CDS, Strasbourg Astronomical Observatory, France.

I have used the GnuDataLanguage \citep{GDL2010, GDL2011, GDL2022} and the Fawlty Language (\url{http://www.flxpert.hu/fl/}) in my analysis.

\bibliography{murthy}{}

\begin{thebibliography}{}
\makeatletter
\relax
\def\mn@urlcharsother{\let\do\@makeother \do\$\do\&\do\#\do\^\do\_\do\%\do\~}
\def\mn@doi{\begingroup\mn@urlcharsother \@ifnextchar [ {\mn@doi@} {\mn@doi@[]}}
\def\mn@doi@[#1]#2{\def\@tempa{#1}\ifx\@tempa\@empty \href {http://dx.doi.org/#2} {doi:#2}\else \href {http://dx.doi.org/#2} {#1}\fi \endgroup}
\def\mn@eprint#1#2{\mn@eprint@#1:#2::\@nil}
\def\mn@eprint@arXiv#1{\href {http://arxiv.org/abs/#1} {{\tt arXiv:#1}}}
\def\mn@eprint@dblp#1{\href {http://dblp.uni-trier.de/rec/bibtex/#1.xml} {dblp:#1}}
\def\mn@eprint@#1:#2:#3:#4\@nil{\def\@tempa {#1}\def\@tempb {#2}\def\@tempc {#3}\ifx \@tempc \@empty \let \@tempc \@tempb \let \@tempb \@tempa \fi \ifx \@tempb \@empty \def\@tempb {arXiv}\fi \@ifundefined {mn@eprint@\@tempb}{\@tempb:\@tempc}{\expandafter \expandafter \csname mn@eprint@\@tempb\endcsname \expandafter{\@tempc}}}

\bibitem[\protect\citeauthoryear{{Akshaya}, {Murthy}, {Ravichandran}, {Henry}  \& {Overduin}}{{Akshaya} et~al.}{2018}]{Akshaya2018}
{Akshaya} M.~S.,  {Murthy} J.,  {Ravichandran} S.,  {Henry} R.~C.,   {Overduin} J.,  2018, \mn@doi [\apj] {10.3847/1538-4357/aabcb9}, \href {https://ui.adsabs.harvard.edu/abs/2018ApJ...858..101A} {858, 101}

\bibitem[\protect\citeauthoryear{{Akshaya}, {Murthy}, {Ravichandran}, {Henry}  \& {Overduin}}{{Akshaya} et~al.}{2019}]{Akshaya2019}
{Akshaya} M.~S.,  {Murthy} J.,  {Ravichandran} S.,  {Henry} R.~C.,   {Overduin} J.,  2019, \mn@doi [\mnras] {10.1093/mnras/stz2186}, \href {https://ui.adsabs.harvard.edu/abs/2019MNRAS.489.1120A} {489, 1120}

\bibitem[\protect\citeauthoryear{{Anderson}, {Henry}, {Brune}, {Feldman}  \& {Fastie}}{{Anderson} et~al.}{1979}]{Anderson1979}
{Anderson} R.~C.,  {Henry} R.~C.,  {Brune} W.~H.,  {Feldman} P.~D.,   {Fastie} W.~G.,  1979, \mn@doi [\apj] {10.1086/157510}, \href {http://adsabs.harvard.edu/abs/1979ApJ...234..415A} {234, 415}

\bibitem[\protect\citeauthoryear{{Boissier} et~al.,}{{Boissier} et~al.}{2015}]{Boissier2015}
{Boissier} S.,  et~al., 2015, \mn@doi [\aap] {10.1051/0004-6361/201526089}, \href {http://adsabs.harvard.edu/abs/2015A%26A...579A..29B} {579, A29}

\bibitem[\protect\citeauthoryear{{Bowyer}}{{Bowyer}}{1991}]{Bowyer1991}
{Bowyer} S.,  1991, \mn@doi [\araa] {10.1146/annurev.aa.29.090191.000423}, \href {http://adsabs.harvard.edu/abs/1991ARA%26A..29...59B} {29, 59}

\bibitem[\protect\citeauthoryear{{Castelli} \& {Kurucz}}{{Castelli} \& {Kurucz}}{2004}]{Castelli2004}
{Castelli} F.,  {Kurucz} R.~L.,  2004, ArXiv Astrophysics e-prints, \href {http://adsabs.harvard.edu/abs/2004astro.ph..5087C} {0405087}

\bibitem[\protect\citeauthoryear{{Chiang}, {M{\'e}nard}  \& {Schiminovich}}{{Chiang} et~al.}{2019}]{Chiang2019}
{Chiang} Y.-K.,  {M{\'e}nard} B.,   {Schiminovich} D.,  2019, \mn@doi [\apj] {10.3847/1538-4357/ab1b35}, \href {https://ui.adsabs.harvard.edu/abs/2019ApJ...877..150C} {877, 150}

\bibitem[\protect\citeauthoryear{{Coulais} et~al.,}{{Coulais} et~al.}{2010}]{GDL2010}
{Coulais} A.,  et~al., 2010, in {Mizumoto} Y.,  {Morita} K.-I.,   {Ohishi} M.,  eds,  Astronomical Society of the Pacific Conference Series Vol. 434, Astronomical Data Analysis Software and Systems XIX. p.~187

\bibitem[\protect\citeauthoryear{Coulais et~al.,}{Coulais et~al.}{2011}]{GDL2011}
Coulais A.,  et~al., 2011, arXiv preprint arXiv:1101.0679

\bibitem[\protect\citeauthoryear{{Draine}}{{Draine}}{2003}]{Draine_scat2003}
{Draine} B.~T.,  2003, \mn@doi [\apj] {10.1086/379118}, \href {http://adsabs.harvard.edu/abs/2003ApJ...598.1017D} {598, 1017}

\bibitem[\protect\citeauthoryear{{Driver} et~al.,}{{Driver} et~al.}{2016}]{Driver2016}
{Driver} S.~P.,  et~al., 2016, \mn@doi [\apj] {10.3847/0004-637X/827/2/108}, \href {http://adsabs.harvard.edu/abs/2016ApJ...827..108D} {827, 108}

\bibitem[\protect\citeauthoryear{{Feldman}, {Brune}  \& {Henry}}{{Feldman} et~al.}{1981}]{Feldman_hotgas1981}
{Feldman} P.~D.,  {Brune} W.~H.,   {Henry} R.~C.,  1981, \mn@doi [\apjl] {10.1086/183657}, \href {https://ui.adsabs.harvard.edu/abs/1981ApJ...249L..51F} {249, L51}

\bibitem[\protect\citeauthoryear{{Green}, {Schlafly}, {Zucker}, {Speagle}  \& {Finkbeiner}}{{Green} et~al.}{2019}]{Green2019}
{Green} G.~M.,  {Schlafly} E.,  {Zucker} C.,  {Speagle} J.~S.,   {Finkbeiner} D.,  2019, \mn@doi [\apj] {10.3847/1538-4357/ab5362}, \href {https://ui.adsabs.harvard.edu/abs/2019ApJ...887...93G} {887, 93}

\bibitem[\protect\citeauthoryear{{Hamden}, {Schiminovich}  \& {Seibert}}{{Hamden} et~al.}{2013}]{Hamden2013}
{Hamden} E.~T.,  {Schiminovich} D.,   {Seibert} M.,  2013, \mn@doi [\apj] {10.1088/0004-637X/779/2/180}, \href {http://adsabs.harvard.edu/abs/2013ApJ...779..180H} {779, 180}

\bibitem[\protect\citeauthoryear{{Henry}}{{Henry}}{1977}]{Henry_radiation_field1977}
{Henry} R.~C.,  1977, \mn@doi [\apjs] {10.1086/190436}, \href {http://adsabs.harvard.edu/abs/1977ApJS...33..451H} {33, 451}

\bibitem[\protect\citeauthoryear{{Henry}}{{Henry}}{1991}]{Henry1991}
{Henry} R.~C.,  1991, \mn@doi [\araa] {10.1146/annurev.aa.29.090191.000513}, \href {http://adsabs.harvard.edu/abs/1991ARA%26A..29...89H} {29, 89}

\bibitem[\protect\citeauthoryear{{Henry} \& {Murthy}}{{Henry} \& {Murthy}}{1993}]{Henry1993}
{Henry} R.~C.,  {Murthy} J.,  1993, \mn@doi [\apjl] {10.1086/187105}, \href {http://adsabs.harvard.edu/abs/1993ApJ...418L..17H} {418, L17}

\bibitem[\protect\citeauthoryear{{Henry}, {Feldman}, {Fastie}  \& {Weinstein}}{{Henry} et~al.}{1978}]{Henry_ngp1978}
{Henry} R.~C.,  {Feldman} P.~D.,  {Fastie} W.~G.,   {Weinstein} A.,  1978, \mn@doi [\apj] {10.1086/156278}, \href {https://ui.adsabs.harvard.edu/abs/1978ApJ...223..437H} {223, 437}

\bibitem[\protect\citeauthoryear{{Henry}, {Murthy}, {Overduin}  \& {Tyler}}{{Henry} et~al.}{2015}]{Henry2015}
{Henry} R.~C.,  {Murthy} J.,  {Overduin} J.,   {Tyler} J.,  2015, \mn@doi [\apj] {10.1088/0004-637X/798/1/14}, \href {http://adsabs.harvard.edu/abs/2015ApJ...798...14H} {798, 14}

\bibitem[\protect\citeauthoryear{{Hensley} \& {Draine}}{{Hensley} \& {Draine}}{2023}]{Hensley2023}
{Hensley} B.~S.,  {Draine} B.~T.,  2023, \mn@doi [\apj] {10.3847/1538-4357/acc4c2}, \href {https://ui.adsabs.harvard.edu/abs/2023ApJ...948...55H} {948, 55}

\bibitem[\protect\citeauthoryear{{Henyey} \& {Greenstein}}{{Henyey} \& {Greenstein}}{1941}]{Henyey1941}
{Henyey} L.~G.,  {Greenstein} J.~L.,  1941, \mn@doi [\apj] {10.1086/144246}, \href {http://adsabs.harvard.edu/abs/1941ApJ....93...70H} {93, 70}

\bibitem[\protect\citeauthoryear{{Holberg}}{{Holberg}}{1986}]{Holberg1986}
{Holberg} J.~B.,  1986, \mn@doi [\apj] {10.1086/164834}, \href {http://adsabs.harvard.edu/abs/1986ApJ...311..969H} {311, 969}

\bibitem[\protect\citeauthoryear{{Jo}, {Seon}, {Min}, {Edelstein}  \& {Han}}{{Jo} et~al.}{2017}]{Jo2017_h2}
{Jo} Y.-S.,  {Seon} K.-I.,  {Min} K.-W.,  {Edelstein} J.,   {Han} W.,  2017, \mn@doi [The Astrophysical Journal Supplement Series] {10.3847/1538-4365/aa8091}, \href {https://ui.adsabs.harvard.edu/abs/2017ApJS..231...21J} {231, 21}

\bibitem[\protect\citeauthoryear{{Jura}}{{Jura}}{1979}]{Jura1979}
{Jura} M.,  1979, \mn@doi [\apj] {10.1086/156788}, \href {http://adsabs.harvard.edu/abs/1979ApJ...227..798J} {227, 798}

\bibitem[\protect\citeauthoryear{{Kulkarni}}{{Kulkarni}}{2022}]{Kulkarni2022}
{Kulkarni} S.~R.,  2022, \mn@doi [\pasp] {10.1088/1538-3873/ac689e}, \href {https://ui.adsabs.harvard.edu/abs/2022PASP..134h4302K} {134, 084302}

\bibitem[\protect\citeauthoryear{{Lenz}, {Hensley}  \& {Dor{\'e}}}{{Lenz} et~al.}{2017}]{Lenz2017}
{Lenz} D.,  {Hensley} B.~S.,   {Dor{\'e}} O.,  2017, \mn@doi [\apj] {10.3847/1538-4357/aa84af}, \href {https://ui.adsabs.harvard.edu/abs/2017ApJ...846...38L} {846, 38}

\bibitem[\protect\citeauthoryear{{Markkanen}}{{Markkanen}}{1979}]{Markkanen1979}
{Markkanen} T.,  1979, \aap, \href {http://adsabs.harvard.edu/abs/1979A%26A....74..201M} {74, 201}

\bibitem[\protect\citeauthoryear{{Marshall}, {Robin}, {Reyl{\'e}}, {Schultheis}  \& {Picaud}}{{Marshall} et~al.}{2006}]{Marshall2006}
{Marshall} D.~J.,  {Robin} A.~C.,  {Reyl{\'e}} C.,  {Schultheis} M.,   {Picaud} S.,  2006, \mn@doi [\aap] {10.1051/0004-6361:20053842}, \href {http://adsabs.harvard.edu/abs/2006A%26A...453..635M} {453, 635}

\bibitem[\protect\citeauthoryear{{Martin} \& {Bowyer}}{{Martin} \& {Bowyer}}{1990}]{Martin1990_lines}
{Martin} C.,  {Bowyer} S.,  1990, \mn@doi [\apj] {10.1086/168376}, \href {https://ui.adsabs.harvard.edu/abs/1990ApJ...350..242M} {350, 242}

\bibitem[\protect\citeauthoryear{{Martin}, {Hurwitz}  \& {Bowyer}}{{Martin} et~al.}{1990}]{Martin1990_h2}
{Martin} C.,  {Hurwitz} M.,   {Bowyer} S.,  1990, \mn@doi [\apj] {10.1086/168681}, \href {https://ui.adsabs.harvard.edu/abs/1990ApJ...354..220M} {354, 220}

\bibitem[\protect\citeauthoryear{{Martin} et~al.,}{{Martin} et~al.}{2005}]{Martin2005}
{Martin} D.~C.,  et~al., 2005, \mn@doi [\apjl] {10.1086/426387}, \href {http://adsabs.harvard.edu/abs/2005ApJ...619L...1M} {619, L1}

\bibitem[\protect\citeauthoryear{{Mathis}, {Whitney}  \& {Wood}}{{Mathis} et~al.}{2002}]{Mathis2002}
{Mathis} J.~S.,  {Whitney} B.~A.,   {Wood} K.,  2002, \mn@doi [\apj] {10.1086/341007}, \href {http://adsabs.harvard.edu/abs/2002ApJ...574..812M} {574, 812}

\bibitem[\protect\citeauthoryear{{Morrissey} et~al.,}{{Morrissey} et~al.}{2007}]{Morrissey2007}
{Morrissey} P.,  et~al., 2007, \mn@doi [\apjs] {10.1086/520512}, \href {http://adsabs.harvard.edu/abs/2007ApJS..173..682M} {173, 682}

\bibitem[\protect\citeauthoryear{{Murthy}}{{Murthy}}{2009}]{Murthyreview2009}
{Murthy} J.,  2009, \mn@doi [\apss] {10.1007/s10509-008-9855-y}, \href {http://adsabs.harvard.edu/abs/2009Ap%26SS.320...21M} {320, 21}

\bibitem[\protect\citeauthoryear{{Murthy}}{{Murthy}}{2014}]{Murthy2014apj}
{Murthy} J.,  2014, \mn@doi [\apjs] {10.1088/0067-0049/213/2/32}, \href {http://adsabs.harvard.edu/abs/2014ApJS..213...32M} {213, 32}

\bibitem[\protect\citeauthoryear{{Murthy}}{{Murthy}}{2016}]{Murthy_dustmodel2016}
{Murthy} J.,  2016, \mn@doi [\mnras] {10.1093/mnras/stw755}, \href {http://adsabs.harvard.edu/abs/2016MNRAS.459.1710M} {459, 1710}

\bibitem[\protect\citeauthoryear{{Murthy} \& {Henry}}{{Murthy} \& {Henry}}{1995}]{Murthy_model1995}
{Murthy} J.,  {Henry} R.~C.,  1995, \mn@doi [\apj] {10.1086/176012}, \href {http://adsabs.harvard.edu/abs/1995ApJ...448..848M} {448, 848}

\bibitem[\protect\citeauthoryear{{Murthy}, {Hall}, {Earl}, {Henry}  \& {Holberg}}{{Murthy} et~al.}{1999}]{Murthy_voy}
{Murthy} J.,  {Hall} D.,  {Earl} M.,  {Henry} R.~C.,   {Holberg} J.~B.,  1999, \mn@doi [\apj] {10.1086/307652}, \href {http://adsabs.harvard.edu/abs/1999ApJ...522..904M} {522, 904}

\bibitem[\protect\citeauthoryear{{Murthy}, {Henry}  \& {Sujatha}}{{Murthy} et~al.}{2010}]{Murthy_galex_data2010}
{Murthy} J.,  {Henry} R.~C.,   {Sujatha} N.~V.,  2010, \mn@doi [\apj] {10.1088/0004-637X/724/2/1389}, \href {http://adsabs.harvard.edu/abs/2010ApJ...724.1389M} {724, 1389}

\bibitem[\protect\citeauthoryear{{Murthy} et~al.,}{{Murthy} et~al.}{2025}]{Murthy2025_alice}
{Murthy} J.,  et~al., 2025, \mn@doi [\aj] {10.3847/1538-3881/ada4a4}, \href {https://ui.adsabs.harvard.edu/abs/2025AJ....169..103M} {169, 103}

\bibitem[\protect\citeauthoryear{{Onaka} \& {Kodaira}}{{Onaka} \& {Kodaira}}{1991}]{Onaka1991}
{Onaka} T.,  {Kodaira} K.,  1991, \mn@doi [\apj] {10.1086/170526}, \href {http://adsabs.harvard.edu/abs/1991ApJ...379..532O} {379, 532}

\bibitem[\protect\citeauthoryear{Park et~al.,}{Park et~al.}{2022}]{GDL2022}
Park J.,  et~al., 2022, \mn@doi [Journal of Open Source Software] {10.21105/joss.04633}, 7, 4633

\bibitem[\protect\citeauthoryear{{Perryman} et~al.,}{{Perryman} et~al.}{1997}]{Perryman1997}
{Perryman} M.~A.~C.,  et~al., 1997, \aap, \href {http://adsabs.harvard.edu/abs/1997A%26A...323L..49P} {323}

\bibitem[\protect\citeauthoryear{{Planck Collaboration} et~al.,}{{Planck Collaboration} et~al.}{2016}]{PlanckDust2016}
{Planck Collaboration} et~al., 2016, \mn@doi [\aap] {10.1051/0004-6361/201424945}, \href {http://adsabs.harvard.edu/abs/2016A%26A...586A.132P} {586, A132}

\bibitem[\protect\citeauthoryear{{Porras-Bedmar}, {Meyer}  \& {Horns}}{{Porras-Bedmar} et~al.}{2024}]{PorrasBedmar2024_axion}
{Porras-Bedmar} S.,  {Meyer} M.,   {Horns} D.,  2024, \mn@doi [\prd] {10.1103/PhysRevD.110.103501}, \href {https://ui.adsabs.harvard.edu/abs/2024PhRvD.110j3501P} {110, 103501}

\bibitem[\protect\citeauthoryear{{Puthiyaveettil}, {Murthy}  \& {Fix}}{{Puthiyaveettil} et~al.}{2010}]{Puthiyaveettil2010}
{Puthiyaveettil} S.,  {Murthy} J.,   {Fix} J.~D.,  2010, \mn@doi [\mnras] {10.1111/j.1365-2966.2010.17149.x}, \href {http://adsabs.harvard.edu/abs/2010MNRAS.408...53P} {408, 53}

\bibitem[\protect\citeauthoryear{{Reynolds}}{{Reynolds}}{1990}]{Reynolds1990}
{Reynolds} R.~J.,  1990, in {Bowyer} S.,  {Leinert} C.,  eds,  IAU Symposium Vol. 139, The Galactic and Extragalactic Background Radiation. p.~157

\bibitem[\protect\citeauthoryear{{Schiminovich}, {Friedman}, {Martin}  \& {Morrissey}}{{Schiminovich} et~al.}{2001}]{Schiminovich2001}
{Schiminovich} D.,  {Friedman} P.~G.,  {Martin} C.,   {Morrissey} P.~F.,  2001, \mn@doi [\apjl] {10.1086/338656}, \href {http://adsabs.harvard.edu/abs/2001ApJ...563L.161S} {563, L161}

\bibitem[\protect\citeauthoryear{{Schlegel}, {Finkbeiner}  \& {Davis}}{{Schlegel} et~al.}{1998}]{Schlegel1998}
{Schlegel} D.~J.,  {Finkbeiner} D.~P.,   {Davis} M.,  1998, \mn@doi [\apj] {10.1086/305772}, \href {http://adsabs.harvard.edu/abs/1998ApJ...500..525S} {500, 525}

\bibitem[\protect\citeauthoryear{{Seon} et~al.,}{{Seon} et~al.}{2011}]{Seon2011}
{Seon} K.-I.,  et~al., 2011, \mn@doi [\apjs] {10.1088/0067-0049/196/2/15}, \href {http://adsabs.harvard.edu/abs/2011ApJS..196...15S} {196, 15}

\bibitem[\protect\citeauthoryear{{Shelton} et~al.,}{{Shelton} et~al.}{2001}]{Shelton2001}
{Shelton} R.~L.,  et~al., 2001, \mn@doi [\apj] {10.1086/322478}, \href {https://ui.adsabs.harvard.edu/abs/2001ApJ...560..730S} {560, 730}

\bibitem[\protect\citeauthoryear{{Welsh}, {Lallement}, {Vergely}  \& {Raimond}}{{Welsh} et~al.}{2010}]{Welsh2010}
{Welsh} B.~Y.,  {Lallement} R.,  {Vergely} J.-L.,   {Raimond} S.,  2010, \mn@doi [\aap] {10.1051/0004-6361/200913202}, \href {http://adsabs.harvard.edu/abs/2010A%26A...510A..54W} {510, A54}

\bibitem[\protect\citeauthoryear{{Witt} \& {Petersohn}}{{Witt} \& {Petersohn}}{1994}]{Witt1994}
{Witt} A.~N.,  {Petersohn} J.~K.,  1994, in {Cutri} R.~M.,  {Latter} W.~B.,  eds,  Astronomical Society of the Pacific Conference Series Vol. 58, The First Symposium on the Infrared Cirrus and Diffuse Interstellar Clouds. p.~91

\bibitem[\protect\citeauthoryear{{Witt}, {Friedmann}  \& {Sasseen}}{{Witt} et~al.}{1997}]{Witt1997}
{Witt} A.~N.,  {Friedmann} B.~C.,   {Sasseen} T.~P.,  1997, \apj, \href {http://adsabs.harvard.edu/abs/1997ApJ...481..809W} {481, 809}

\bibitem[\protect\citeauthoryear{Ysard, Koehler, Jones, Miville-Desch{\^e}nes, Abergel  \& Fanciullo}{Ysard et~al.}{2015}]{Ysard2015}
Ysard N.,  Koehler M.,  Jones A.~P.,  Miville-Desch{\^e}nes M.-A.,  Abergel A.,   Fanciullo L.,  2015, Astronomy and Astrophysics, 577

\makeatother
\end{thebibliography}
\bibliographystyle{mnras}

\end{document}